\documentclass[prd,aps,twocolumn]{revtex4-1}
\usepackage{epsfig,amsmath,amssymb}
\usepackage{mathrsfs}
\usepackage[usenames,dvipsnames]{color}
\usepackage{hyperref}

\begin{document}
\title{Effective General Relativistic Description of Jamming in Granular Matter }

\author{Soumendra Kishore Roy}
\email{soumendrakishoreroy09@gmail.com}
\affiliation{Department of Physics, State University of New York at Stony Brook, Stony Brook, NY 11794 , USA.}

\author{Pratyusava Baral}      \email{baralpratyusava@gmail.com}
\affiliation{Department  of  Physics, University of Wisconsin-Milwaukee,  Milwaukee, WI 53201, USA.}

\author{Ratna Koley}
\email{ratna.physics@presiuniv.ac.in}
\affiliation{Department of Physics, Presidency University, Kolkata 700073, India.}

\author{Parthasarathi Majumdar}
\email{bhpartha@gmail.com}
\affiliation{School of Physical Sciences, Indian Association for the Cultivation of Science, Kolkata 700032, India.} 

\begin{abstract}

We propose here that certain observational features of granular matter in the infrared limit, exhibiting the phenomenon of {\it jamming}, arise from an underlying effective general relativistic description. The proposal stems from the assumption (which we justify on physical grounds) that grains in granular matter move freely in an {\it effective} curved Riemannian space. The termination of their trajectories at the onset of jamming is obtained from the focussing of a converging congruence of geodesics in such a space, as a solution of the Raychaudhuri equation for such congruences. This may happen irrespective of whether or not the curvature is sourced by external stresses (via an effective Einstein equation), although the properties of the resultant jammed state solution do differ in the two cases. A definite prediction of this geometrical approach is the negative role played by those trajectories which twist about each other, in reaching the jammed state. The local symmetries of granular interaction, translational and rotational invariance (corresponding to `force balance' and `torque balance' in standard force-based approaches to jamming) are inherent in the effective general relativity framework. A recently-proposed effective elasticity model of the jammed state, based on a tensorial variant of standard electrostatics (Vector Charge Theory), is seen to be entirely subsumed within the linearized version of the effective general relativistic description.
      
\end{abstract}

\maketitle

\section{Introduction} 

Granular matter, and indeed many other
athermal amorphous materials, exhibit an intriguing property known as
{\it jamming}. This property manifests in abrupt termination of trajectories of test particles of granular matter (grains), like in a dense solid, under certain specific conditions. On relaxing these conditions, grains may `flow' freely, as streamlines of a fluid. This dichotomic behaviour is at the heart of difficulties in the extensive ongoing and past assays in  theoretical analyses of the laws that
govern such materials \cite{andreo13}, \cite{beh-chak2018}. The laws appear to have firm underpinnings in out-of-equilibrium, nonlinear physical interactions \cite{andreo13}, even though jamming per se appears to be a {\it steady state} feature \cite{beh-chak2018}. For such materials, there is an absence of a clear separation of scales between the microphysics and macrophysics. Despite this,  `critical behaviour' similar to thermal critical phenomena, characterized by a correlation length that diverges at the jamming-unjamming transition
\cite{linag98}, and power law critical exponents, have been observed in such materials. In addition, time-series analyses of density heterogeneities in granular matter under shear stress exhibit appearance of
filamentary structures in the density profile \cite{beh-chak2018}. 

\begin{figure}[h].
\centering
\includegraphics[width=5cm,height=4.5cm]{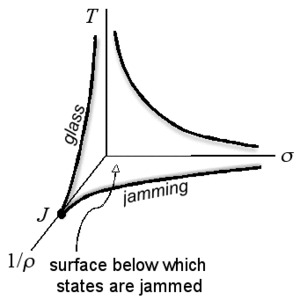}
\label{Fig.1}
\caption {Jamming `phase' diagram, adapted from ref. \cite{linag98} }
\end{figure}

One early characterization of this phenomenon is in terms of strong stress heterogeneities occuring within the material, with and without an externally applied stress. From this standpoint, spatial density fluctuations observed in a granular material like sand, are attributed to force chains \cite{cates98} - force ordered alignments of grains which inhibit motion of test grains along the alignment direction (friction). Thus, in
each such alignment direction, these materials behave like a solid, while, in other directions, the material may indeed ‘flow’, akin to a fluid. Possible force chain configurations are restricted by imposing equilibrium with respect to net force and torque, implying an underlying translational and rotational symmetry in the granular matter. Externally applied stresses may actually break these symmetries, leading to appearance of density inhomogeneities, even though the force-chain model does not appear to provide a {\it natural}, detailed description of how such inhomogeneities may arise.
 
A recent alternative to the force-chain description is an effective elasticity model \cite{bulbulprl2020}, where a tensorial version of electrostatics known as Vector Charge Theory (VCT) is proposed to describe the infrared behaviour of jammed states. Correlators of the tensorial `electric field' are computed and shown to exhibit a behavior which seems to exhibit the density inhomogeneities observed in shear-induced jamming. The main problem with this approach is that it is inherently a static and linear theory, which may provide a picture of the jamming state, but fails to illuminate the dynamical aspects of jamming phenomena. The model does not immediately generalize to a nonlinear, dynamical theory, and usual Maxwell electrodynamics provides no easy guidance, since it cannot accommodate vectorial charges.    

The great variety of observed and simulated phenomena pertaining to
granular matter motivates attempts to construct a simple, {\it unified},
effective description in the infrared limit, which captures the essence of some of the more important  
phenomenological properties, without depending too strongly on the underlying microscopics.

The phenomenon of jamming, as already mentioned, is characterized by the termination of test grain trajectories under certain high density conditions of granular matter. This would imply that these trajectories must converge {\it prior} to reaching the jammed state, i.e., they would tend to {\it focus}, as they evolved in time. Such focussing of particle trajectories can be analyzed geometrically, without recourse to local forces, through the time evolution of the spatial variation of the velocity vector field of a {\it congruence} of grain trajectories, assuming that the trajectories correspond to free particle motion in an effective {\it curved} Riemannian space. This assumption enables the use of the formalism of the Raychaudhuri equation \cite{akr55, poisson04} involving the focusing or otherwise of geodesic congruences in Riemannian manifolds. The Ricci curvature of the space is a key element of the Raychaudhuri equation. If one now makes the additional assumption that the Ricci curvature is related to the external matter stresses causing the curvature, the focusing of the trajectories can then be related to these stresses. This is tantamount to the assumption that the effective Riemannian spacetime arises from Einstein's equation of general relativity. Thus, our proposal is to analyze the phenomenon of jamming in terms of an {\it effective general relativistic description} which is inherently geometrical, while being nonlinear and dynamical, properties that are suggested from observed aspects of granular matter. The proposal appears all the more logical in the context of granular matter where applied shear stresses play a crucial role in jamming, if one recalls that material stresses, in addition to mass (energy and momentum) density, are natural sources of curvature in general relativity. 

Indeed, in absence of external stresses, a critical density (packing fraction) of granular matter, is necessary for jamming to occur. Such a jammed state is often seen to be {\it isotropic} when the underlying geometry has a spherical isometry. With appropriate external shear stresses present, smaller packing fractions than the critical threshold can still lead to a jamming solution, with appearance of additional features like anisotropic density inhomogeneities. Both these aspects of granular matter manifest naturally within the proposed effective general relativity description, as we shall show in the sequel.  The appearance of density inhomogeneities in presence of external stresses, as a departure from isotropic jamming, is realized in the effective general relativistic description within the theory of {\it cosmological perturbations}, reversed in time. 

A novel feature, not explored extensively in the granular matter literature to the best of our knowledge, is the effect of {\it twist} or {\it rotation} of grain trajectories around neighbouring ones. We shall show that the Raychaudhuri equation {\it predicts} that this behaviour of grain trajectories acts against realizing the jamming state, it is an {\it anti-jamming} feature which may not be too hard to experimentally probe. In other words, with external stresses present alongwith rotation, larger stresses may be needed to reach the jamming state compared to the situation in absence of rotation of trajectories about each other. If this phenomenon is indeed a novel one, as we suspect it is, it is a direct test of the effective general relativity approach to jamming in granular matter. 

We mention in passing that the imposition of local force and torque balance, required within the force-chain description mentioned earlier, are {\it inherent} in the effective general relativistic description, as local translational and rotational invariance. No additional requirements need be imposed on the geometry. General relativity is  well-known to be inherently non-linear \cite{Straumann}, and to admit a great many {\it time-dependent} solutions (like gravitational collapse solutions) which, from a statistical standpoint correspond to distinctly {\it non-equilibrium} processes. In many cases though, {\it constant-time} spatial slices do reveal steady state aspects of such solutions. This underlies the correspondence proposed in this paper.

The paper is organized as follows : in section 2, mathematical details of focusing phenomena of test grain trajectories are presented within the formalism of the Raychaudhuri equation. An important consequence of this equation is the role of twist of rotation on the evolution and possible covergence of grain trajectories, which we present as a unique prediction of the proposed framework.  In section 3, we deal with isotropic jamming phenomena in granular matter, within the effective general relativistic correspondence, which relates it to a Schwarzschild black hole. Properties of this effective black hole are seen to reproduce observed phenomena of isotropic jamming related to divergence of correlation length at the jamming-unjamming transition. The next section (4) discusses the situation with application of external (shear) stresses whereby isotropicity is lost, while jamming is seen to lead to appearance of {\it filaments} of density inhomogeneities observed in shear-stressed granular matter. Appealing to the theory of time-reversed cosmological perturbations, the effective general relativistic description is seen to lead, to lowest order in the perturbations, density heterogeneities in the matter. While the broad features seem consistent, a comparison between detailed features of the heterogeneities for sheared granular matter, and those in the effective general relativity description, has not been attempted here, and is left to a subsequent publication. In the next section (5), the Vector Charge Theory and its main consequences are derived from a {\it linearized} approximation of our effective general relativistic description, showing that our proposal has a greater efficacy over the one it subsumes. We end in section 6 with a discussion on the scope of our approach, and also the pending tasks before it.

\section{Focusing of test grain trajectories}

Test grain trajectories are to taken as free particle paths within an ambient curved space, each being labelled parametrically by ${\vec x} = {\vec x}(t)$. We deal here with a {\it congruence} of such trajectories, i.e., a non-intersecting collection of grain paths such that at any time $t$, only one path passes through any point in space. The collection of velocities ${\vec u} \equiv d{\vec x}/dt$ at any $t$ thus constitute a {\it field} in space, whose spatial variations may be studied. We define the Jacobi field $B_{ij} \equiv \nabla_{i} u_{j}~,~i,j=1,2,3$ as measuring such a variation, where $\nabla_{i}$ is not just a partial derivative, but a {\it covariant} derivative necessary in curved space. Usually, when a tangent vector to a curved space is naively parallel transported to a neighboring point, it ceases to be a tangent vector at the new point. It is thus necessary to project the naively transported vector down to the tangent space at the new point, to yield what may be {\it defined} as the actual parallelly transported tangent vector. A covariant derivative compliments naive parallel transport effected by a partial derivative, by means of the affine connection which captures precisely this {\it projection} to the tangent space. 

The Jacobi field $B_{ij}$ is thus a second rank tensor field in the ambient curved space; as such it admits the decomposition into its irreducible components, i,e, its traceless symmetric, trace and antisymmetric pieces :
\begin{eqnarray}
B_{ij} &=& \sigma_{ij} + \frac13 h_{ij} \theta + \omega_{ij } \label{irr} \\
\sigma_{ij} &\equiv & \nabla_{i} u_{j} + \nabla_{j} u_{i} -\frac13 h_{ij} \theta \label{sher} \\
\theta & \equiv & h_{ij} \nabla_{i} u_{j} \label{expn} \\
\omega_{ij} &\equiv & \nabla_{i} u_{j} - \nabla_{j} u_{i} \label{twst}  \\
\nonumber
\end{eqnarray}
where $h_{ij}$ is the metric of the three dimensional space at any fixed time slice $t$. in the equations above, $\omega_{ij}$ arises from grain trajectories twisting around each other, and is thus called the {\it twist} or the {\it rotation} tensor, $\theta$ arises from the fractional cross-sectional volume expansion/contraction at each $t$ and is called the {\it expansion} while $\sigma_{ij}$ corresponds to the deformation in the cross-sectional surface at every $t$ due to differential speeds of grains, and is called the {\it shear}. We attempt to represent the time evolution of these geometrical quantities visually in Fig.2 where the light blue spheres depict cross-sectional surfaces of the congruence of trajectories at every $t$, with time running downwards in Fig.2. This time evolution is described by the Raychaudhuri equation \cite{akr55}, \cite{poisson04},, which involves the {\it spacetime} completion of the above tensors,
\begin{eqnarray}
\frac{d\theta}{dt} = \omega^2 - \frac13 \theta^2 - \sigma^2 - R_{\mu \nu} u^{\mu} u^{\nu} \label{rceq}
\end{eqnarray}
with $\mu, \nu=0,1,2,3$ and $R_{\mu \nu}$ being the Ricci curvature tensor of the effective curved spacetime. Two key ideas underlie this equation : first of all is our basic assumption that grain trajectories are basically free particle paths (geodesics) in an ambient curved space and second, the curvature of that space is defined by the change in a vector field (like the velocity field) when parallel transported around a closed curve in the space. Following ref. \cite{wald}, a short derivation of this equation is sketched in the appendix for a full pseudo-Riemannian spacetime, for the benefit of interested readers.   
\begin{figure}[h].
\centering
\includegraphics[width=5cm,height=4.5cm]{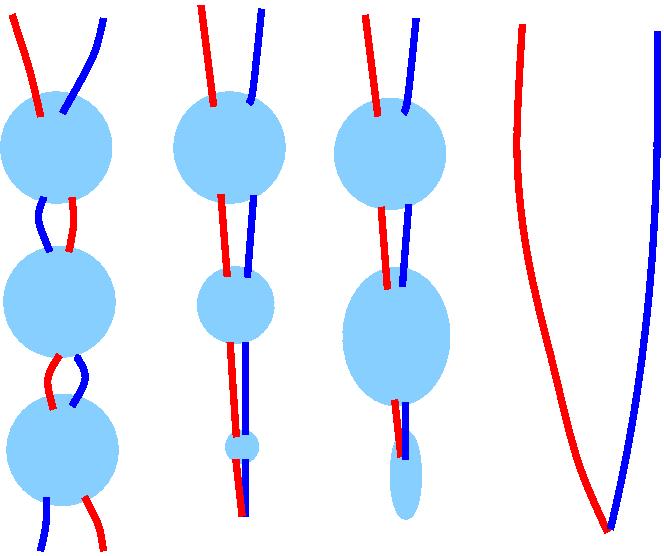}
\label{Fig.2}
\caption {Time evolution of squares of irreducible tensor components of the Jacobi field as they appear in the Raychaudhuri equation : The diagrams from left depict the twist or rotation, (negative) expansion, shear deformation and external curvature causing convergence, respetively. Light blue spheres denote the cross-sectional surface of the paths at each constant time slice $t$. Time flows vertically downward. }
\end{figure} 

As is apparent from eqn (\ref{rceq}) as also from Fig.1, apart from the rotation term, all other terms are negative, since the squares of the irreducible tensors are all positive \cite{wald}, \cite{poisson04}. This is particularly important for convergence of the trajectories as a solution of eqn (\ref{rceq}). Consider the simplest case when the trajectories are free of rotation and shear deformation and the ambient space is Ricci flat. In this case, eqn (\ref{rceq}) reduces to $d\theta/dt = -(1/3) \theta^2$. This equation can be easily integrated between $t=0$ to $t$, yielding the solution
\begin{eqnarray}
\theta^{-1}(t) = \theta^{-1}(0) + 3t \label{sol}
\end{eqnarray}
If the initial value of the expansion parameter, $\theta(0) < 0$, then there is a specific time $t=t_c = -(1/3) \theta^{-1}(0)$ such that the right hand side vanishes. In this case, $\theta(t_c) \rightarrow -\infty$ which signifies a formation of {\it caustics} due to covergence of the congruence of paths. Evolution of the trajectories into the future beyond $t=t_c$ is plagued by ambiguities. This configuration is identified with the jamming state. The expansion parameter signifies the rate of change of {\it fractional volume} or {\it packing fraction} : $\theta \equiv (\delta V)^{-1} (d (\delta V))/dt$. The packing fraction is dynamical, i.e., a function of time, in the onset of jamming situation under consideration. Thus, the initial condition that the expansion parameter in the Raychaudhuri equation  is negative, translates into the packing fraction of granular matter beginning to exceed a critical threshold value, $f_c$ say, as the condition for jamming to occur {\it spontaneously}, i.e., without external stresses present. This behaviour, which follows from the solution of the Raychaudhuri equation in the effective general relativistic description of jamming, reproduces the zero-temperature, zero-external stress behavior of packing fraction (inverse density) for jamming situations depicted in Fig. 1 \cite{linag98}. 

Note that Ricci flatness in general relativity does not imply a flat (i.e., non-curved) ambient space, but rather corresponds to a {\it vacuum} solution of Einstein equation with vanishing external stress-energy tensor. The simplest spacetime in this case is the spherically symmetric (isotropic) Schwarzschild spacetime, which, in this effective framework, models {\it isotropic} jamming. We shall consider this situation in somewhat more detail in the next section. 

We end this section with two important observations : first of all, with rotation and shear of grain trajectories still vanishing, if the ambient space is driven by an external stress-energy tensor, we have in eqn (\ref{rceq}), the Ricci curvature term on the left-hand side replaced by $(8\pi G/c^4)(T_{\mu \nu} - (1/2) g_{\mu \nu} T)u^{\mu} u^{\nu}$ via the effective general relativistic Einstein equation, where $T_{\mu \nu}$ is the external stress-energy tensor, $T$ its spacetime trace with the metric of the effective curved spacetime $g_{\mu \nu}$. If the stress-energy tensor, contracted with the 4-velocities $u^{\mu} u^{\nu}$, is positive, i.e., it corresponds (say in the rest frame of any particular trajectory) to positive energy density of the external source ($T_{00} > 0$), then this term continues to have a negative sign in (\ref{rceq}). This then contributes to making the expansion negative, thus leading to convergence of the congruence of paths, i.e., to jamming. Depending upon how big the external stresses are, it is more than likely, that even if the initial value of the expansion parameter is {\it not negative}, i.e., the packing fraction is suboptimal and by itself this would not lead to jamming, large enough external stresses will nevertheless drive the system to convergence of grain trajectories implying caustic formation which, in turn, signifies jamming. The phenomenon of jamming due to external shear stresses in granular matter, with suboptimal packing fractions, has been extensively discussed in the literature on granular matter \cite{maj-beh05}, \cite{beh-chak2011}, \cite{beh-chak2018}. This property follows in the proposed effective framework from the Raychaudhuri equation. 

Another consequence of eqn (\ref{rceq}) is perhaps not so well-understood, namely the effect of {\it rotation} or twist of grain paths around each other. It is clear that for non-vanishing rotation, eqn (\ref{rceq}) {\it predicts} that the convergence of paths may be {\it inhibited}, because of the positive sign of the squared rotation term in the equation. Thus, rotation would lead to partial {\it unjamming}, according to this idea. To the best of our knowledge, we do not know if experiments have actually observed this phenomenon. If not, this is a definite prediction of this effective general relativity approach : if granular matter is made to follow paths that twist around each other, then it would require a larger external shear stress to reach a jamming state, in comparison to the situation of vanishing rotation. How much larger the external stress has to be, of course depends on the strength of the squared rotation term in the equation, which could be a control parameter in a suitably setup experiment. 
        
This {\it anti-jamming} aspect of rotation of grain trajectories is not entirely surprising, if we recall that vortex motion in fluids inherently involves rotation of velocity fields. There is thus a sharp distinction, provided by the Raychaudhuri equation, between those aspects that favour the realization of a jammed state, corresponding to solid-like behaviour of granular matter,  and those which oppose such a realization and favour instead a more fluid-like behaviour. We do not know of any other theoretical framework proposed in the jamming context, where such a sharp distinction is available within the same framework for granular matter.

\section { Isotropically Jammed states \cite{o'hearn2003} :} 

A simple representative of an {\it isotropically} jammed state in a granular substance, in contrast to an {\it unjammed} state is depicted in Fig. 3, showing that for the same total number of grains, jamming clearly entails a {\it contraction} of the spatial volume.   
\begin{figure}[h].
\centering
\includegraphics[width=5cm,height=3.5cm]{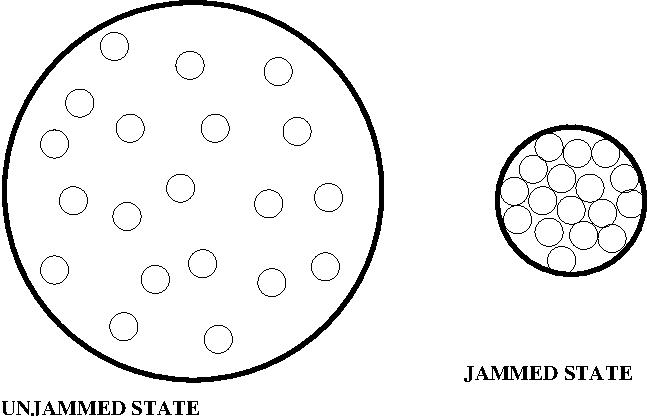}
\label{Fig.3}
\caption { Unjammed vs jammed state, requiring volume contraction}
\end{figure}
In the effective general relativity description of granular material, we propose that jamming is an effective gravitational collapse phenomenon, where the volume contracton of cross-sectional surfaces depicted in Fig.2 under the convergence of evolving grain paths, realizes the phenomenon of formation of {\it trapped} surfaces \cite{penrose65}. In astropysical phenomena like formation of black holes from dying massive stars, trapped surfaces form late in the collapse process, only when a certain threshold density is reached. In the case of granular matter, this is analogous to the critical packing fraction required for formaton of jammed states. In Fig. 4, untrapped and trapped surfaces are shown, with volume expansion of {\it constant-time} (spacelike hypersurface) slices of flat Minkowski spacetime, and volume {\it contraction} occurring for a gravitationally collapsing body. 
\begin{figure}[h] 
\centering
 \includegraphics[width=5cm,height=3.5cm]{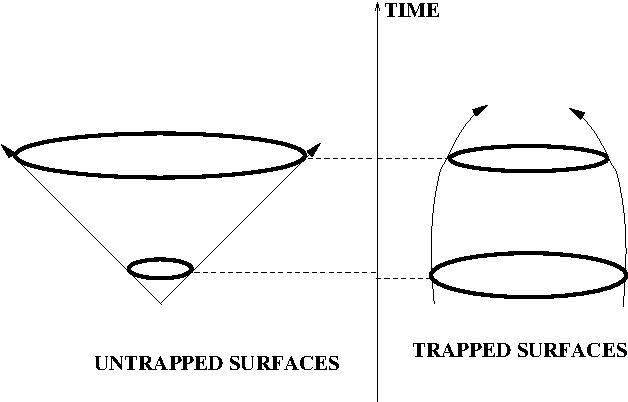}
 \label{Fig.4}
\caption { Untrapped vs trapped surfaces, requiring volume contraction}
\end{figure}
The inevitability of formation of trapped surfaces in gravitational collapse phenomena in astrophysics is shown \cite{penrose65} to be a consequence of the Raychaudhuri equation.

Isotropic jamming corresponds clearly to a situation with spherical symmetry. With $T_{\mu \nu}=0$, the effective {\it vacuum} Einstein equation reduces to $R_{\mu \nu}=0$ which  admits (uniquely) the Schwarzschild solution
\begin{eqnarray}
ds^2 &=& \left( 1 - \frac{r_s}{r} \right) (dx^0)^2 - \left( 1 - \frac{r_s}{r} \right)^{-1} dr^2 \nonumber \\
&-& r^2(d\theta^2 + \sin^2 \theta d\phi^2)  \label{sch}
\end{eqnarray} 
The length scale $r_S$ characterizing this solution is called the Schwarzschild radius;  $r=r_S$ represents the boundary between grains accessible to observers at infinity, and those that are confined to the region $r \leq r_S$ - the {\it effective black hole} region - and hence not similarly accessible. Indeed, jamming as a phenomenon sharply distinguishes between grains that are confined, and grains that are free to move. So the description of the boundary of a jammed state as an effective {\it black hole horizon} in our general relativity framework, is perhaps logical. 

There is also an interchange between $r$  and $x^0=ct$  in (\ref{sch}) for $r < r_S$, the coordinate $r$ plays the role of time, while $x^0$ is a spatial (radial) coordinate. Thus, the {\it interior} solution is a {\it contracting} solution, and it is this solution whose constant time (spatial)  slices are trapped surfaces. A gravitationally collapsing spherical ball of pressureless dust \cite{oppensny39} in astrophysics admits a very similar interior solution once trapped surfaces are formed. The horizon radius $r_S$ is taken to correspond to the length scale characterizing the jammed state.  

The spatial part of the effective black hole metric in eqn (\ref{sch}) permits one to define the squared {\it correlation length} of the jammed region : $dl^2 = (1-r_S/r)^{-1} dr^2 + r^2(d\theta^2 + \sin^2 \theta d\phi^2)$. This correlation length clearly  diverges as $r\rightarrow r_S$, i,e., as one approaches the jamming-unjamming boundary. In statistical mechanical studies of the jamming-unjamming transition \cite{linag98}, a similar phenomenon of divergence of correlation length has been observed at the transition, with the packing fraction serving as an effective order parameter in absence of external stresses. This is a bit mysterious from the viewpoint of thermal phase transitions, given that granular matter is inherently athermal. However, this is apparently quite natural in the geometrical framework being proposed here. An open question at this point is whether the framework can also reproduce the `critical exponents' discerned in the statistical mechanical work \cite{linag98}.   

Since this divergence manifests only in the `Schwarzschild frame' of the effective general relativity description, and not necessarily in other coordinate frames, general coordinate invariance is clearly not a hallmark within the proposed framework. This is not a genuine lacuna, since despite this feature, acoustic and optical analogues of general relativity appear to work well, with robust observational support. However, the spatial symmetries of granular interactions, translational and rotational invariance, are manifest in every frame of the effective general relativity approach.

A simple alternative model to creating jammed states with external stress-energy is the {\it perfect fluid} $T_{\mu \nu} = (P + \rho) u_{\mu} u_{\nu} - P g_{\mu \nu}$, with $P$ and $\rho$ being respectively the pressure and density of the fluid, and $u^{\mu}$ the 4-velocity of the fluid at each point. In absence of external stresses, $P=0$, the interior solution is often approximated in the effective general relativity approach, as a {\it contracting} Friedmann-Lemaitre-Robertson-Walker (FLRW) spacetime \cite{poisson04}, expressed in the frame comoving with the contracting time-slice, as,
\begin{eqnarray}
ds^2 = a^2(t) (dt^2 - d{\vec x}^2)~.~t \rightarrow {\rm conformal ~time} \label{flrw}
\end{eqnarray}
We have set $c=1$ for convenience. As $t$ increases, the {\it scale factor} $a(t)$ decreases, exhibiting a contracting spacetime. As contraction proceeds, the granular matter begins to form spherical jammed surfaces with a horizon as an envelope. On every spatial slice. the metric is proportional to  the flat Euclidean metric. This is another manner in which formation of trapped surfaces in astrophysical processes involving gravitational collapse have an obvious parallel with formation of jammed states in granular matter.

\section {Jamming due to external Stress perturbations} 

When $P \neq 0$ a richer variety of density fluctuations appear, under externally applied stress, especially a shear stress \cite{maj-beh05}, \cite{beh-chak2018}, as seen in Fig. 5 below.
\begin{figure}[h] 
\centering
 \includegraphics[width=6cm,height=3cm]{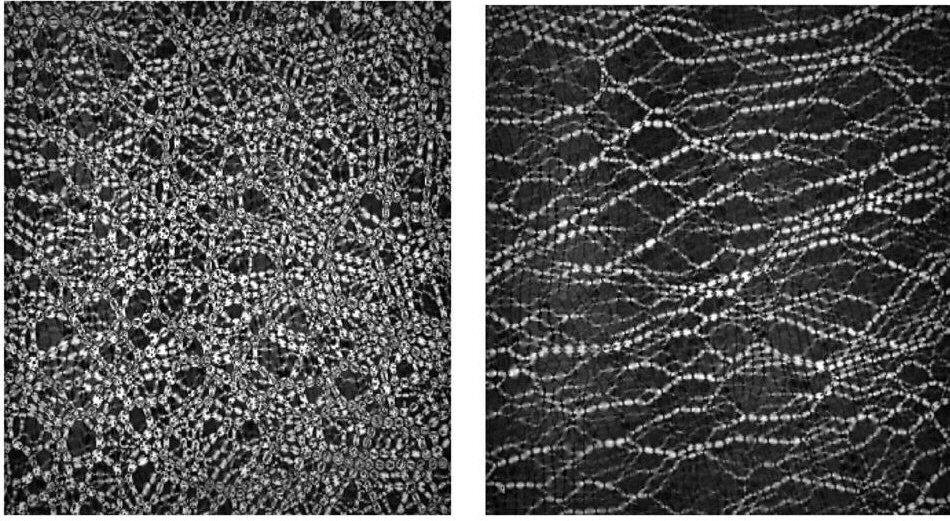} 
 \label{Fig.5}
\caption { Photoelastic images of Isotropic jammed and shear-jammed states (adapted from ref. \cite{maj-beh05})}
\end{figure}
In the effective general relativity framework, this is implemented by the introduction of linear perturbations around the isotropic, homogeneous unperturbed {\it unjammed} state, corresponding to the initial rate of change of fractional volume being non-negative. These linear perturbations which are not restricted to be isotropic or homogeneous, give rise to structures in the density profile of granular substances. In the comoving time frame, these perturbations are given by $g_{\mu\nu} = a^2(t) (\eta_{\mu \nu} + h_{\mu \nu})$, where $\eta_{\mu \nu}$ is the metric of Minkowski spacetime. The metric fluctuations $h_{\mu \nu}$ are subject to {\it gauge} transformations $\delta_{\xi} h_{\mu \nu} = \nabla^{(0)}_{\mu} \xi_{\nu} + \nabla^{(0)}_{\nu} \xi_{\mu}$, where $\xi_{\mu}$ is a 4-vector local gauge parameter, and the covariant gradients are taken with respect to the unperturbed analogue geometry. A set of gauge {\it invariant} potentials $(\Psi, \Phi, \Phi _i, h_{ij}^{\text{TT}})$ \cite{Bardeen1981} (Latin indices running over 1,2,3 belong to 3-space), formed by linear combinations of $h_{\mu \nu}$ components, are our key physical variables. The two scalar modes $\Psi$ and $\Phi$, relevant for cosmological observations in the physical universe, play an important role in the effective general relativity approach to granular matter. We focus on the effect of external stress perturbations on the  changes in the effective spatial geometry. Thus, the density perturbation is given by $\rho' = \rho + {\tilde \rho}$, while the velocity perturbation is given by  $u'_i = u_i + V_i$. The spatial distribution of the gauge invariant perturbation for a given type of matter fluctuation is given by the perturbed Einstein equation restricted to a spatial slice,
\begin{equation}\label{Perturb}
    \nabla^2 \Psi = \frac{\kappa}{2}a^2 \rho {\tilde \rho}~,~ \nabla^2 \Phi _i=-2\kappa \rho a^2(1+w)V_i 
\end{equation}
along with the evolution equations for them \cite{Bardeen1981}, \cite{Peter}. Now, let us define a symmetric second rank tensor $\chi_{ij}$ as,
\begin{equation}\label{Stress}
    \chi_{ij}=\frac{1}{2}(\partial_i\Phi_j+\partial_j\Phi_i)+\partial_i\partial_j\Psi
\end{equation}
$\chi_{ij}$ is automatically curl-free $\epsilon_{ikl}\epsilon_{jmn}\partial^k\partial_m\chi^{ln}=0$, with trace $\chi_{ii}=\nabla^2\Psi=-4\pi a^2 \rho {\tilde \rho}$ which is non-zero for the physical universe. In the physical universe, the vector mode dies out fast due to the expanding spacetime \cite{Peter}, \cite{Mukhanov}. In the analogue gravity/cosmology situation addressed here, even though we are dealing with a {\it contracting} spacetime, we are at liberty to ignore the vector mode in the first place without altering the curl-free and non zero trace condition of $\chi_{ij}$. Thus, the divergence of $\chi_{ij}$ is given by,
\begin{equation}\label{Gauss}
    \partial_i \chi_{ij}= - 4 \pi \rho a^2 \partial_j {\tilde \rho}
\end{equation}
The static limit of the above equation \eqref{Gauss}, with the source being a Gaussian random variable, simulates the mechanical response of an amorphous solid with the identification $\chi_{ij} \leftrightarrow \sigma_{ij}$ and $-4\pi \rho a^2\partial_j {\tilde \rho} \leftrightarrow f_j$. where $\sigma_{ij}$ is  the  jamming stress of granular matter and $f_j$ is the applied force. The scale factor $a$ on a spatial slice remains a free parameter depending on the particular granular matter under consideration. Numerical solution of the resulting Poisson's equation in the static limit leads to the following graphical representation of density fluctuation of granular matter under external density and pressure perturbations. We show that an isotropically jammed state caused by an isotropic external stress can be explained using the Poisson equation. We start with a $10,000  \times 10,000$ uniform two dimensional grid which is synonymous to a homogeneous unperturbed granular medium. Perturbations as uniform random numbers centered around the mean density are introduced and the potential is allowed to evolve using a simple finite difference partial differential equation solver. The resulting potential represents an isotropically jammed  state with density fluctuations as shown in the figure. Our simulation closely resembles  the images of photoelastic material observed in \cite{maj-beh05} in case of isotropic jamming. The white signifies an overdense region while the black region signifies a density lower than the mean value.

This is depicted in Fig. 6, with the unperturbed granular density assumed to be uniform for simplicity.
\begin{figure}[h]
\begin{center}
\includegraphics[width=7cm]{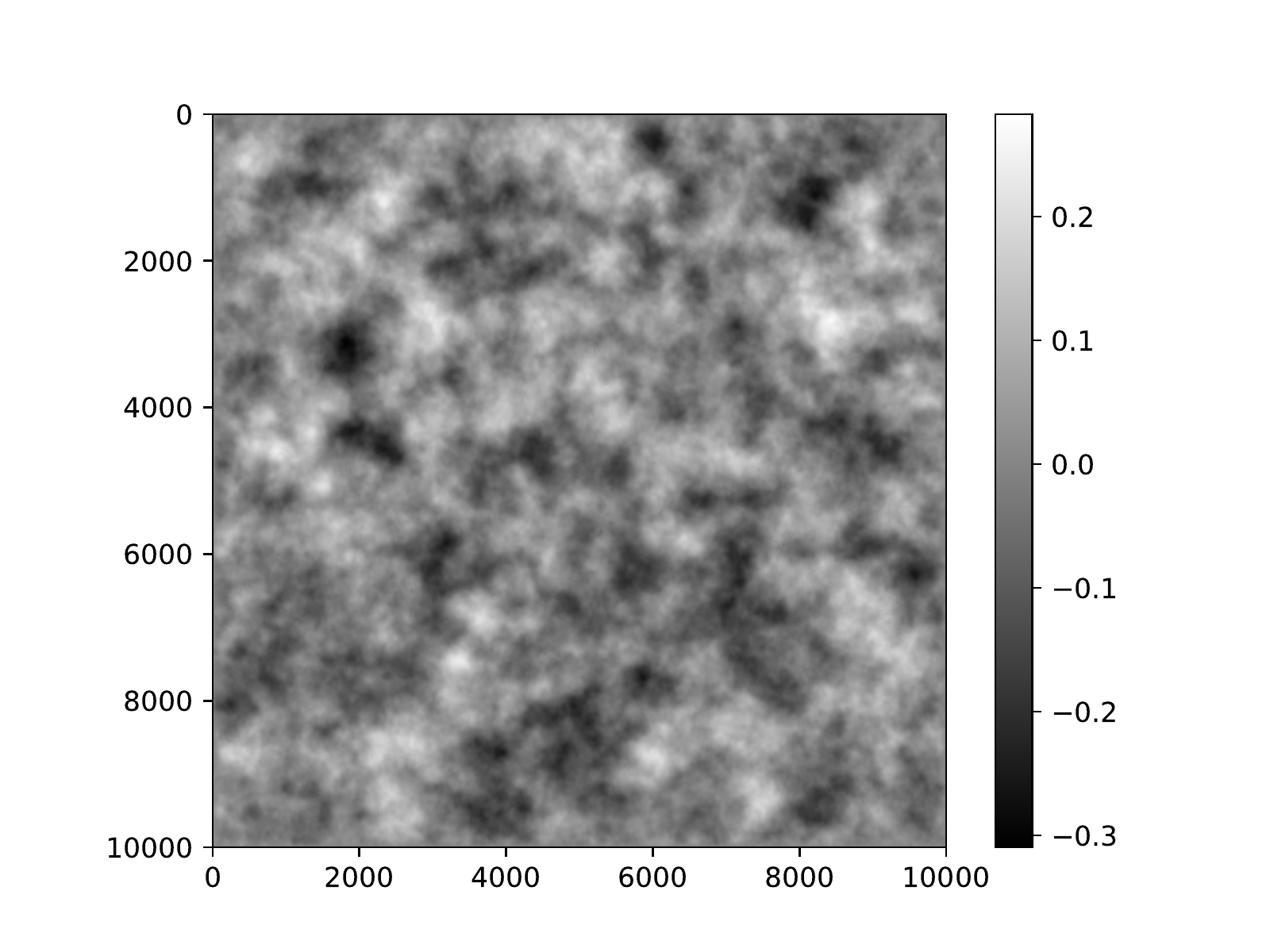}
\caption{Granular matter surface density fluctuations on a spatial slice induced by external isotropic stress perturbations. The unperturbed density profile is assumed to be uniform for simplicity}
\label{Fig.6}
\end{center}
\end{figure}

A further interesting aspect of this correspondence is the analogue lensing in a photo-elastic material. The equation \eqref{Perturb} gives the freedom to choose the minimum value of $\Psi$ to be $0$. The corresponding perturbed metric $ds^2=(1+2\Psi)dt^2-(1-2\Psi)\delta_{ij}dx^i dx^j$ shows the lensing with refractive index $n=1-2\Psi$ \cite{Straumann}. This fact can be simulated in a photo-elastic material under external stress. The physical measurable quantities in this scenario are the correlators of refractive index which depend on the central difference rather than on an absolute measure.

\section {Vector Charge Theory from Linearized Effective General Relativity} 

Recently, an emergent continuum elasticity model named as Vector Charge Theory (VCT), attempting to capture the infrared behaviour of granular matter has been proposed \cite{bulbulprl2020}, based on {\it fracton} gauge theory \cite{pret}. The notion of a {\it vector} charge is untenable in Maxwell electrodynamics, so the tensorial generalization attempted in VCT is somewhat contrived. However, we demonstrate in what follows that the VCT is entirely {\it subsumed} within the {\it linearized} approximation of the general relativistic correspondence proposed here, in the sense that its physical consequences can be derived from this correspondence. The linearized approximation is necessitated by the fact that the VCT is a linear theory. Unlike general relativity which is inherently nonlinear, the nonlinear version of VCT has not been formulated yet. Furthermore, material stresses provide a natural source of spacetime curvature in general relativity, so when applied to granular matter, physical consequences like jamming ensue straightforwardly, without ad hoc added features.  

The corresponding spacetime metric $g_{\mu \nu} = \eta_{\mu \nu} + h_{\mu \nu}$ where, recall that $\eta_{\mu \nu}$ is the Minkowski metric. Invariance of the linearized Einstein equation (retaining only first degree terms in $h_{\mu \nu}$) under the linearlized gauge transformations $\delta_{\xi} h_{\mu \nu} = \partial_{\mu} \xi_{\nu} + \partial_{\nu} \xi_{\mu}$, permits us to construct the relevant gauge invariant fluctuation degrees of freedom $\Theta,~ \Phi (3-scalars),~ \Sigma_i (transverse~3-vector), ~h^{TT}_{ij} (traceless ~symmetric~ tensor)$. The stress-energy tensor $T_{ij}$, on a constant-time  slice, obeys the conservation laws $\partial_i T_{ij} = 0~,~T_{ij}=T_{ji}$ corresponding respectively to spatial translational and rotational symmetry, as a leftover from global Poincar\'e symmetry in the full analogue spacetime. The full $T_{\mu \nu}$ is decomposed into the 3-scalars $\rho, ~P$, 3-vector (transverse) $S_i$ and traceless symmetric tensor ${\bar \sigma}_{ij}$. In the static linearized limit, Einstein equation reduces to
\begin{eqnarray}
\nabla^2 \Theta = - 8 \pi \rho & , & \nabla^2 \Phi = 4 \pi (\rho +3P) \nonumber \\
\nabla^2 \Sigma_i = - 16 \pi S_i & , & \nabla^2 h_{ij}^{TT} = 16 \pi {\bar \sigma}_{ij} \label{linee}
\end{eqnarray} 
The radiative mode $h_{ij}^{TT}$ and its source ${\bar \sigma}_{ij}$ are ignored for the purpose of establishing this correspondence.

Define a tensorial electric field $E_{ij}$ by
\begin{eqnarray}
E_{ij} \equiv \frac12(\partial_i \Sigma_j + \partial_j \Sigma_i) + \partial_i \partial_j \Phi~. \label{corr}
\end{eqnarray}
Clearly, $E_{ij} = E_{ji}~,~ E_{ii} \neq 0$, and its physicality is guaranteed by the gauge invariance of the constituent variables $\Sigma_i$ and $\Phi$. Using the linearized Einstein equations (\ref{linee}), it is easy to verify that
\begin{eqnarray}
\partial_i E_{ij} = -8\pi (S_i + \partial_i \rho)  &\equiv& 4\pi \rho_i  \nonumber \\
\epsilon_{ikl} \epsilon_{jmn} \partial_k \partial_m E_{ln} &=& 0~, \label{tenses}
\end{eqnarray}
so that, the vector charge density is also well-defined in terms of components of the stress-energy tensor of our analogue gravity description. Equation (\ref{tenses}) are the fundamental equations for the tensorial electrostatics  proposed in the emergent elasticity model of ref. \cite{bulbulprl2020}, based on fracton gauge theory \cite{pret}. 

The gauge invariant fluctuations $\Theta,~ \Phi,~ \Sigma_i, ~h^{TT}_{ij}$, all space and time components of $h_{\mu \nu}^T$, can be most easily derived from $h_{\mu \nu}$ through the projection operator \cite{anarya2019}, \cite{soum2020} $\Pi_{\mu \nu \rho \lambda} : h_{\mu \nu}^T =\Pi_{\mu \nu}^{\rho \lambda} h_{\rho \lambda}$, where,
\begin{eqnarray}
\Pi_{\mu \nu}^{\rho \lambda} \equiv \delta_{(\mu}^{\rho} \delta_{\nu)}^{\rho} - \frac{1}{k^2} k_{(\mu}k^{\rho} \delta_{\nu)}^{\lambda} + \frac{1}{k^4} k_{\mu}k_{\nu}k^{\rho}k^{\lambda}. ~\label{proj} 
\end{eqnarray}     
where, brackets in the indices imply symmetrization. Computing the appropriate components for $E_{ij}$ from eqn. (\ref{corr}) with the indices restricted to spatial indices, and taking the infrared limit, the {\it free} correlation function $\langle E_{ij}({\vec k}) E_{kl}(-{\vec k}) \rangle$ is derived to be
\begin{eqnarray}
C_{ijkl}^{(free)} \propto \delta_{(i| k} \delta_{|j) l} - \frac{k_{(i|} k_{ (k|} \delta_{|j) |l)}}{{\vec k}^2} + \frac{k_i k_j k_k k_l}{{\vec k}^4}   
\end{eqnarray}
in agreement with the key eqn. (9) of ref. \cite{bulbulprl2020}. This justifies our claim that the proposed correspondence between granular matter and general relativity, in the linearized approximation, reproduces physical consequences of the VCT.

We should perhaps mention that there may be other definitions of $E_{ij}$ which would  lead to the same correspondence, so that the correspondence may afford extra freedoms of definition. Observe that within this correspondence, it is not necessary to add any tensorial polarization field due to bound vector charges as in the VCT.

\section {Concluding Remarks} 

Analogue gravity in barotropic, inviscid  fluids with acoustic perturbations \cite{unruh81}, \cite{visser99} is now well-understood both theoretically and observationally. In that case, the knowledge of the basic fluid mechanics equations (Euler-Navier-Stokes) for centuries provides a relative advantage over the case of granular matter where fundamental theoretical understanding is still being sought. Hence an a `analogue gravity' description of such materials have to start {\it ab initio}, with little in standard physics to act as a guide. Why should such an approach work at all ? A priori, this cannot be answered rightaway, only the future can tell.

One advantage of our proposal of an effective general relativistic framework to understand the low energy behaviour of granular matter is that it is directly experimentally testable. The role of twisting grain paths in reaching a jammed state, whether this entails a larger external shear stress than when the paths are non-twisting, ought to be accessible to experimental verification in the near future. This opportunity to `falsify' the framework is a robust aspect that it possesses. As discussed in section 2, by embodying both the `solid-like' and `fluid-like' aspects of granular matter within it, the Raychaudhuri equation stands unique among theoretical proposals towards unravelling this difficult frontier of physics.    

Among pending issues is that of {\it repulsive} granular interactions, since gravity is supposed to be inherently attractive. Repulsive granular interactions may emerge from our approach by introducing a {\it cosmological constant} into the effective Einstein theory, or by considering `white hole' geometries, or even analogues of expanding universes. The analogue of `Hubble friction' in that case may help stabilize and stimulate growth of structures in the density  profile. This is  certainly a direction worthy of exploration. 

Perhaps one can also turn the analogy on its head, and use granular matter physics to explore phenomena that are yet unobserved in real cosmology. An example would be the existence of primordial gravitational waves which in the real universe are still shrouded in mystery \cite{Planck}. Similarly, aspects of the CMB spectrum like non-Gaussianity \cite{Mukhanov} may also show up as a feature of granular media observed experimentally. All-in-all, the assay portends an interesting future.   

\noindent{\it Acknowledgments :} This Letter is inspired by a recent webinar on granular matter given by Prof Bulbul Chakraborty, and ensuing discussions. We are very grateful to Prof. Chakraborty for introducing us to various aspects of granular matter, which stimulated our thinking to a large extent. We are also thankful to Dr Suchetana Chatterjee for asking an important question and to Dr Subhro Bhattacharjee for useful correspondence. SKR and
PB would like to thank Mr Satyanu Bhadra for referring us to ref. \cite{andreo13} and for discussions, and to Mr Arindam Das for computational help.    

\appendix*
\section{Sketch of derivation of eqn (\ref{rceq}) \cite{wald}}

The 4-velocity defined by $u^{\mu}, ~\mu = 0,1,2,3$ satisfies the geodesic equation, which basically says that it is the equation for a curve for a {\it free} particle in the ambient curved spacetime, such that its tangent vector at any value of the proper time parameter $\tau$ is parallel transported along itself, 
\begin{eqnarray}
u^{\mu} \nabla_{\mu} u^{\nu} = 0 \label{geod}
\end{eqnarray}
Computing the evolution of the spacetime completion of the Jacobi field $B_{\mu \nu}= \nabla_{\nu} u_{\mu}$, along any geodesic, we obtain
\begin{eqnarray}
u^{\mu} \nabla_{\mu} B_{\nu \rho} &=& u^{\mu} \nabla_{\rho} \nabla_{\mu} u_{nu} + u^{\mu} [\nabla_{\mu}, \nabla_{\rho}] u_{\nu} \nonumber \\
&=& - (\nabla_{\rho} u^{\mu})(\nabla_{\mu} u_{\nu} + u^{\mu} R_{\mu \rho \nu \lambda} u^{\lambda} \nonumber \\
&=& -B^{\mu}_{\rho} B_{\nu \mu} + u^{\mu} R_{\mu \rho \nu \lambda} u^{\lambda} \label{jac}
\end{eqnarray}
In line 2 above we have used eqn (\ref{geod}). Upon tracing (\ref{jac}) with the spatial metric $h_{\mu \nu}$, and using the fact that the irreducible tensors $\omega$, $\sigma$ in (\ref{irr}) in the text and the spatial metric $h$ are mutually orthogonal, eqn (\ref{rceq}) follows.

\end{document}